\documentclass[onecolumn]{aastex63}

\usepackage[utf8]{inputenc}
\usepackage[a4paper, total={6in, 8in}]{geometry}
\usepackage{natbib}
\usepackage{longtable}
\usepackage[page]{appendix}

\usepackage{amsmath}
\usepackage{bm}
\usepackage{mathtools}

\usepackage{graphicx}
\usepackage{esvect}

\def\aap{Astro. Astrophys.}
\def\apjl{Astrophys. J. Let.}

\def\apjs{Astrophys. J., Supp.}

\begin{document}

\shorttitle{Global Photometric Modeling}
\shortauthors{Judkovsky et al.}

\received{July 2, 2023}

\title{The Advantages of Global Photometric Models in Fitting Transit Variations}

\correspondingauthor{Yair Judkovsky}
\author[0000-0003-2295-8183]{Yair Judkovsky}
\affiliation{Weizmann Institute of Science, Rehovot 76100 Israel}
\email{yair.judkovsky@weizmann.ac.il}

\author[0000-0002-9152-5042]{Aviv Ofir}
\affiliation{Weizmann Institute of Science, Rehovot 76100 Israel}

\author[0000-0001-9930-2495]{Oded Aharonson}
\affiliation{Weizmann Institute of Science, Rehovot 76100 Israel}
\affiliation{Planetary Science Institute, Tucson, AZ, 85719-2395 USA}

%\date{November 2019}

\graphicspath{{./}{Figures/}}

%\maketitle

\begin{abstract}
Estimation of planetary orbital and physical parameters from light-curve data relies heavily on the accurate interpretation of Transit Timing Variations (TTV) measurements. In this letter, we review the process of TTV measurement and compare two fitting paradigms - one that relies on making transit-by-transit timing estimates and then fitting a TTV model to the observed timings and one that relies on fitting a global flux model to the entire light-curve data simultaneously. The latter method is achieved either by solving for the underlying planetary motion (often referred to as "photodynamics"), or by using an approximate or empirical shape of the TTV signal. We show that across a large range of the transit SNR regime, the probability distribution function (PDF) of the mid-transit time significantly deviates from a Gaussian, even if the flux errors do distribute normally. Treating the timing uncertainties as if they are distributed normally leads, in such a case, to a wrong interpretation of the TTV measurements. We illustrate these points using numerical experiments and conclude that a fitting process that relies on a global flux fitting rather than the derived TTVs, should be preferred. 
\end{abstract}


\section{Background} \label{sec:Background}
Interpreting photometric light curves is increasingly common in the age of multiple ground-based and space observatories such as Kepler \citep{BoruckiEtAl2010} and TESS \citep{RickerEtAl2010}, which led to the discovery and characterization of thousands of exoplanets.
The transit method provides information regarding the orbital period, phase, and several normalized geometrical properties: the planetary size, semi-major axis, and impact parameter -- all relative to the stellar radius \citep[e.g.][]{SeagerMallen-Ornelas2003}. If the stellar mass is known, the absolute semi-major axis can be inferred, and if the stellar radius is known, the absolute planetary radius can be inferred. The planet's mass, though, cannot be obtained using periodic transits alone. The eccentricity of the orbit is related to the transit duration \citep[e.g. ][]{Carter2008}, but is degenerate with the orbital orientation.
To infer the planetary mass and orbital shape, perturbations among the planets need to exist in the system. In this case, Transit Timing Variations \citep[TTV, ][]{AgolSteffenSariClarkson2005,HolmanMurray2005} arise; this phenomenon was found abundantly in the Kepler population \citep{HolczerEtAl2016, OfirEtAl2018}.
 
Multiple authors have extensively studied the functional form of the TTV signal. Of specific importance was the introduction of the fundamental super-period TTV \citep{LithwickXieWu2012}, which enabled a fast linear inversion of the sine-like TTV signal to masses and eccentricities, up to the degeneracies involved. These degeneracies can be removed if a synodic "chopping" signal is visible \citep{DeckAgol2015} or if second-order TTV modes are visible \citep{HaddenLithwick2016}. Another avenue for using TTVs to infer planetary properties is to use full N-body integrations to predict transit times \citep[e.g.][]{DeckAgolHolmanNesvorny2014}, driven by a non-linear fitting process that compares the predicted times with the time estimates obtained from the data. Such methods were used to interpret the planetary parameters in many systems \citep[][and many more]{HaddenLithwick2014,HaddenLithwick2017,JontoffHutter2014,JontofHutter2016}. 
 
Light-curve data that contains TTVs can be interpreted in two different methods. One method is to perform a global fit on the light-curve photometric data, either by using a dynamical model of the planetary motion \citep[often called photodynamics, e.g.][]{Carter2011, GrimmEtAl2018, Judkovsky2022b}, or by using an empirical TTV functional form \citep[e.g.][]{OfirEtAl2018}. The second method is to estimate the individual properties of each transit, typically the mid-transit times, thus transforming the light-curve flux data into timing data, and then to interpret these to estimate the planetary properties via the measured pattern of the TTVs \citep[e.g.][]{HaddenLithwick2016,JontofHutter2016,HaddenLithwick2017}. 

In this work, we compare the performance of these two methods at various single-transit SNR values. We do not present a full statistical analysis of the errors, but instead, we aim to illustrate the consequences of utilizing typical practices in light-curve and TTV fitting, highlighting the applicable methods and their advantages or shortcomings. We achieve this goal by performing a numerical experiment to show that the inaccurate propagation of the flux errors into transit timing errors can yield wrong results in the TTV analysis. This elaborates on and clarifies the arguments presented in \citet[][\S2.3]{OfirEtAl2018}. In addition, we stress that the following analysis does not aim to show that full photodynamical modeling outperforms a transit-by-transit timing analysis in a particular exoplanetary system or a particular modeling code but rather to show that performing a fit on the photometric data outperforms a transit-by-transit analysis, regardless of the underlying model from which the TTV shape is derived.

\section{Transit Timing from Flux}
A prerequisite for the method that relies on using the individual times of mid-transit is the translation of the light-curve flux data to a set of individual transit time estimates, usually done using a non-linear fitting algorithm. The timing values and errors are then used to fit the TTV model parameters, which are functions of the planetary physical parameters. An underlying assumption within this process is that a Gaussian can approximate the posterior probability density function (PDF) (which, given a uniform prior, is equal to the likelihood of the data) as a function of the transit time. In other words, the assumption is that the propagation of the flux errors to timing errors can be done by summarizing the PDF properties in two numbers: the expectation value and standard deviation. This method was used, for example, in several ground-breaking works dealing with exoplanetary parameter estimation, such as \citet[][and others] {HaddenLithwick2016,HaddenLithwick2017}. See formulation in \citet[][Appendix B.1]{HaddenLithwick2016}. As we show below, this assumption might not hold even for perfectly normally-distributed flux values for a large range of transit SNR values, where the individual transit SNR is defined as
 \begin{equation}
     \rm SNR=\frac{\delta}{\sigma_{\rm F}}\sqrt{N_{\rm pts}}\simeq \frac{(R_{\rm p}/R_*)^2}{\sigma_{\rm F}}\sqrt{N_{\rm pts}},
 \end{equation}
 where $\delta$ is the transit depth, $\sigma_{\rm F}$ is the typical uncertainty on the flux, $N_{\rm pts}$ is the number of data points within a single transit event, and $R_{\rm p}/R_*$ is the planet-to-star radius ratio.

The transit SNR is related to the measurement precision of individual mid-transit time. A deeper transit would have more distinctive ingress and egress, enabling a more accurate timing measurement. The transit timing error is a complicated function of the planetary parameters, and in practice, it is obtained from a non-linear fitting process. We are not aware of a closed-form formula for this error. However, \citet{HolczerEtAl2016} provides a simplified empirical benchmark, which states that the timing error is roughly $100\,{\rm min}/{\rm SNR}$ (with some spread around this value).

As we show below, the PDF of the mid-transit timing is not Gaussian. This deviation grows as the SNR decreases. Adequately propagating the flux errors to timing errors requires not only the mean and standard deviation of the PDF but the entire PDF shape. Propagating the expectation value and standard deviation alone and using them in the TTV fitting as if they represent a Gaussian can yield wrong results, especially when modeling shallow transits. 
 
Assuming that the noise on the flux distributes normally, a global flux fit that integrates the information of all transits simultaneously would not suffer from this problem. In the language of the SNR definition given above, when fitting the entire light curve simultaneously, the effective SNR increases with the square root of the number of transits.
 
Real photometric data may not distribute normally and suffer from red noise, further complicating the light-curve analysis in both methods. Here, we neglect this additional complication and assume that the flux noise distributes normally. We note that if the full PDF of each transit was measured and propagated to the TTV model then, in principle, the transit-by-transit method would have access to the full information needed to produce an equivalent fit to that of the global flux model (up to the appearance of transit variations other than TTV). The issue that makes the transit-by-transit technique problematic is that the PDF of the individual transit time does not distribute normally, even if the flux data does. We demonstrate this argument in the following numerical experiment.

 \section{Numerical Experiment}
In this section, we mimic the process of interpreting a TTV-containing light curve using two methods: In the first method, we estimate the individual transit times and then use them to fit a TTV model; in the second method, we use a global model to fit the light-curve flux directly. Importantly, to isolate these factors we assume perfect knowledge of all other elements, such as the properties of the plant, star, or TTV model shape.

\subsection{Data Generation}
The simulated flux data is generated as follows. First, we calculate a sequence of nominal times of mid-transit for a planet with an orbital period $P=10.13$ days and with a phase that sets the first nominal transit time at $T_{\rm mid,0}=1$ day, along the time span $t=0..1500$ days. We obtain a series of 148 nominal times of mid-transit. Then, we add to these nominal transit times a sinusoidal TTV of the form
\begin{equation}
    \rm TTV=A\cos{\omega t}+B\sin{\omega t},
\end{equation}
where $A=0.01$ day and $B=0.02$ day are constants that together set the amplitude and phase of the TTV, $\omega=0.015 \,{\rm Rad/day}$ is the TTV's angular frequency, and $t$, in this case, is the nominal transit time for which the TTV is calculated.
Adding the TTV to each nominal transit time yields a set of actual transit times. We then assume a circular orbit with a semi-major axis-to-star ratio $a/R_*=15$ and an impact parameter $b=0.3$ and use the obtained orbital geometry to construct a full light-curve using the Mandel-Agol formula \citep{MandelAgol2002} with limb-darkening parameters $u_1=0.36$ and $u_2=0.28$ \citep{Claret2000a}.  All the constants were chosen at values commonly found in Kepler data. This process results in a light curve without any measurement errors. For simplicity, we neglect the effect of binning \citep{Kipping2010a}, although binning must be addressed for real data.

The next step in the flux data generation process is to add noise. We assume a white noise model in which the noise on each data point is distributed as $\mathcal{N}(0,\sigma_{\rm F}^2)$, that is normally with a zero mean and $\sigma_{\rm F}$ standard deviation.

We repeat the experiment several times, varying $\sigma_{\rm F}$ over a wide range. In reality, the noise spectrum is not white, and a red component is present; we elaborate on this later.

\subsection{Data Interpretation}
Having obtained noised light-curve data, we interpret it using the abovementioned methods. The interpretation process aims to find the best-fitting TTV frequency $\omega$ and coefficients $A,\, B$. In the first method, in which the transit times are estimated one by one, we first treat each transit individually. We keep only data points within 0.5 days from the nominal time of mid-transit. We assume perfect knowledge of the planet-to-star radius ratio, the semi-major axis, and the impact parameter. We optimize only for the transit time using the Levenberg-Marquardt (LM) algorithm. This algorithm provides a best-fitting value and an estimated variance on the (only) fitted variable - the mid-transit time. The starting value for the fitted variable is the nominal time of the individual transit under consideration. This starting point should be sufficiently close to the true value to enable convergence because our TTV amplitude is significantly smaller than the transit duration. Having obtained a set of mid-transit times, we subtract from them the linear fit to these times and obtain the simulated measured TTV signal. We then perform a second non-linear fit using the LM algorithm, where this time, the fitted variable is the TTV frequency $\omega$. Because for each $\omega$ value, the best fitting $A, B$ can be found using linear regression, we can optimize the single variable $\omega$ to obtain a best-fitting TTV model. In this case, the starting point for fitting $\omega$ is set to the true underlying $\omega$ value. This would be the case, for example, when the perturbing planet is known because then the TTV super-period can be calculated\citep{LithwickXieWu2012}. This process results in a best-fitting model of the parameters $\omega,\, A,\, B$. We note that for our finite data span, a simultaneous fit for a line and a sinusoid would, in principle, yield a better match than a sequential fit of a line and then a sinusoid; however, given that the super-period is shorter than the data span, the induced error is much smaller than the TTV amplitude.

In the second method, we perform a non-linear fitting for the flux data, simultaneously fitting $\omega,\, A,\, B$ using the non-linear LM algorithm over three variables. This process results in estimates for the values of $A, B, \omega$, and their errors.

\subsection{Results - an Illustrative Example}
\begin{figure*}[h]
    {\includegraphics[width=.9\linewidth]{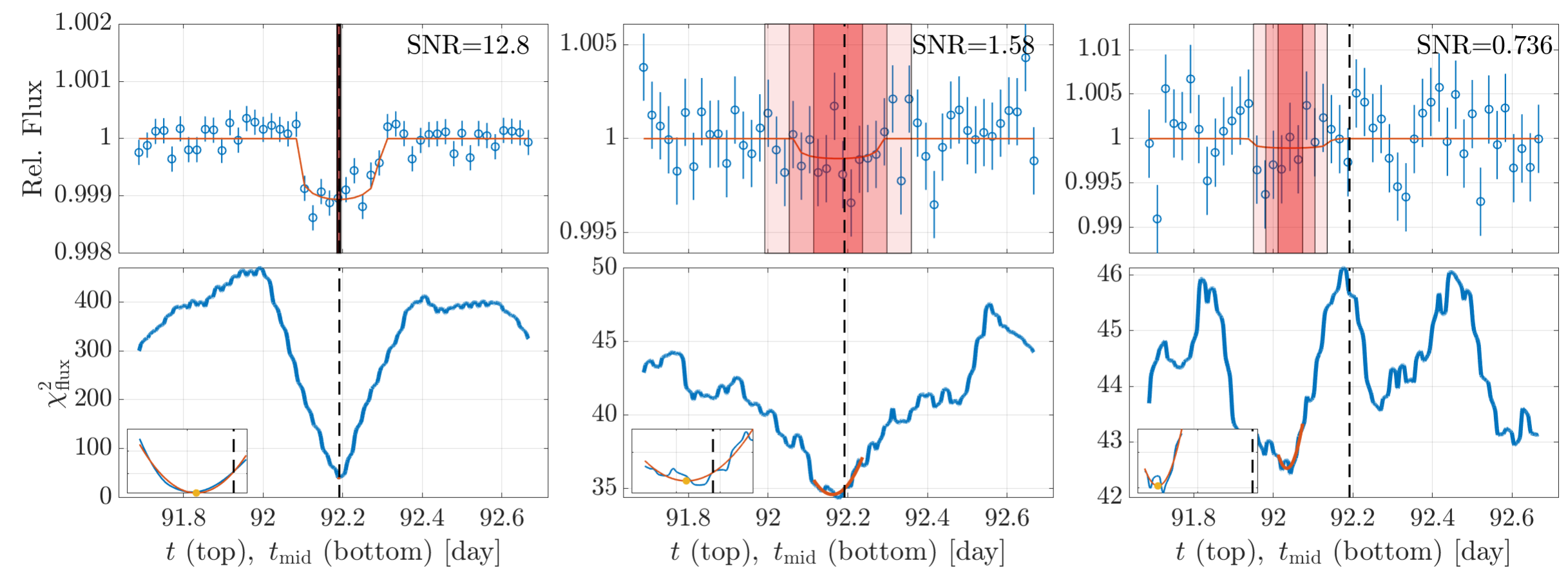}}
    {\caption{Three individual events of different SNR values. Top row: The blue error bars are the synthetic data points, the solid orange line is the best-fitting model, the vertical black dashed line is the true time of mid-transit, and the pink transparent stripes are the 1,2 and $3\,\sigma$ confidence intervals of the time of mid-transit, as retrieved by the LM algorithm (on the left panel they are too narrow to be distinguishable). The black dashed line denotes the true time of mid-transit. Bottom row: For each event, we show the $\chi^2$ surface over different values of time of mid-transit (blue) and a best-fitting parabola for the points within the $1\,\sigma$ confidence interval. The small inserts show a zoom-in on the region where the parabolic fitting was performed, with a circle indicating the parabola vertex. In the high SNR case (left-hand side), there is a clear depression in the $\chi^2$ surface, with a shape similar, but not equal to, a parabola around the best fitting value. As SNR is reduced towards unity (middle), there is still a minimum near the true time of mid-transit, but its shape deviates from a parabola even more than in the high-SNR case. In SNR values lower than unity, the transit shape is erased by the noise (note the relatively shallow vertical span of the entire panel), and other local minima begin to appear in the $\chi^2$ surface, thus leading to a completely wrong estimate of the time of mid-transit.}
    \label{fig:IndividualEvents}}
\end{figure*}

We highlight the results demonstrating the difficulty of propagating flux estimates to time estimates - a limitation arising from the non-Gaussian distribution of the time-of-mid-transit for an individual event.
In Figure~\ref{fig:IndividualEvents}, we show examples of the fit quality of three individual events, each taken from a different numerical experiment with a different single-transit SNR value. This figure demonstrates that for an SNR value significantly larger than unity, there is a clear global minimum in the $\chi^2$ surface which is similar, but not equal to, a parabola, which means that the likelihood distribution is close to, but not exactly, a Gaussian. As the SNR approaches unity from above, the shape of this minimum deviates more and more from a parabola. When the SNR value is smaller than unity, other local minima can appear, overprinting the true transit signal by noise.

\begin{figure*}[h]
    {\includegraphics[width=.9\linewidth]{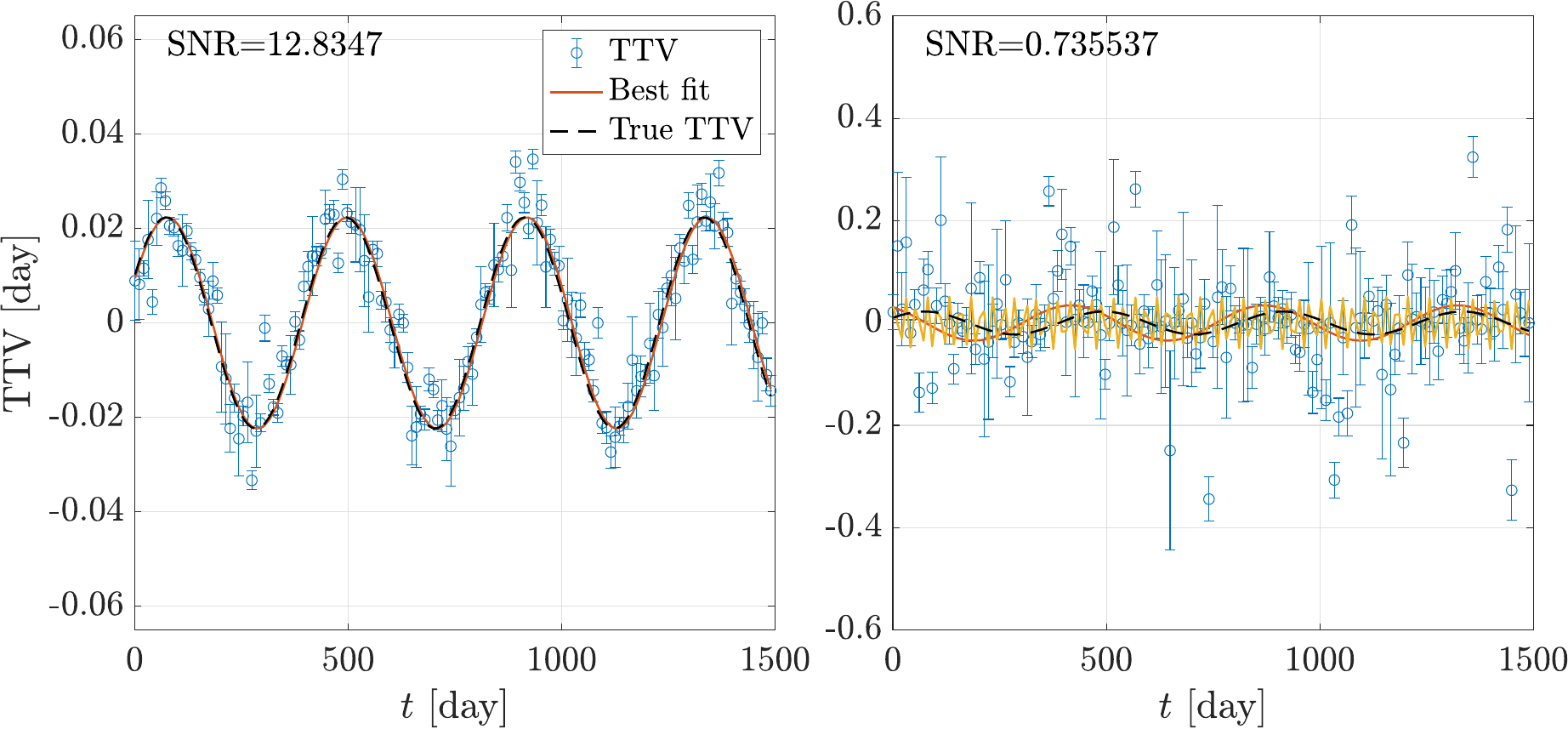}}
    {\caption{The best-fitting TTV solutions for two numerical experiments, each with a different single-transit SNR value. The blue error bars are the TTV estimates from fitting individual transit times, and the solid orange curve is the best-fitting TTV model, obtained from an LM search beginning at the correct $\omega$ value. The yellow curve on the right-hand-side plot is the best-fitting TTV model in the Lomb-Scargle peak frequency. At the high-SNR case (left-hand side), the true TTV frequency is apparent in the estimated TTV obtained from a transit-by-transit analysis, leading to a correct estimation of the TTV frequency $\omega$ up to the estimated error. In the low-SNR case (right-hand side), the noise has erased the transit events, thus leading to wrong or highly uncertain estimates of the individual mid-transit times. The estimated TTV frequency, in this case, is far from the true value.}
    \label{fig:TimingFitQuality}}
\end{figure*}

In Figure~\ref{fig:TimingFitQuality}, we show the fit for the TTV frequency for a high-SNR and a low-SNR case, assuming a normal probability distribution of each individual mid-transit time. The resulting frequency value corresponds to the true TTV frequency in the high-SNR case but significantly differs from the true frequency in the low-SNR case. It is also shown that without prior knowledge of the starting point for $\omega$, a random frequency would be preferred based on a Lomb-Scargle periodogram. This figure demonstrates the end effect of treating the PDF of each individual event timing as a Gaussian: the TTV fit that relies on the individual time estimates and their associated errors may result in a wrong TTV frequency.

\begin{figure*}[h]
    {\includegraphics[width=.9\linewidth]{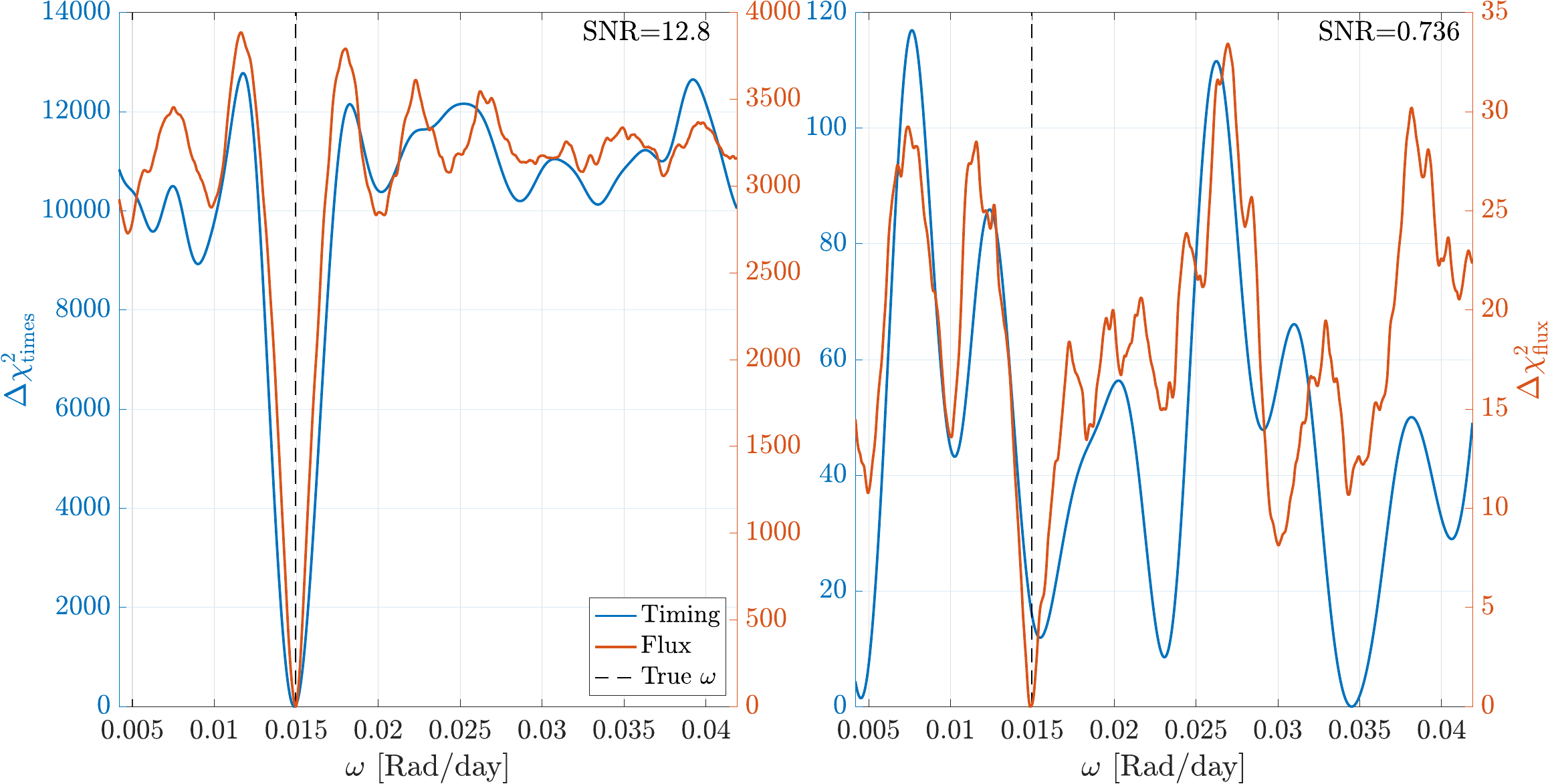}}
    {\caption{The $\chi^2$ surface in the timing method and the flux method for the high-SNR case (left-hand side) and for the low-SNR case (right-hand side). The $\Delta\chi^2$ surface (relative to the minimum) in the method using individual mid-transit times fitting is shown in blue; the $\Delta\chi^2$ surface for the global flux method is shown in orange. A dashed black line indicates the true TTV frequency. In the high-SNR case, both methods generate a clear global deep minimum around the true TTV frequency. In the low-SNR case, additional, deeper minima are found around other frequencies in the timing method. Local minima exist in the flux fitting method, but the global minimum remains around the true TTV frequency.}
    \label{fig:Chi2SurfaceFluxVsTiming}}
\end{figure*}

In Figure~\ref{fig:Chi2SurfaceFluxVsTiming}, we show the $\Delta \chi^2$ surface as a function of  the TTV frequency $\omega$ for both methods: fitting individual mid-transit times and fitting a global model over the flux. Along this surface, the parameters $A,\, B$ are kept constant at their true values. Both methods yield a clear global minimum around the true TTV frequency in the high-SNR case. The transit-by-transit method generates a wider minimum and generally a different likelihood surface, explained by the incorrect error propagation of the individual timing estimates. In the low-SNR case, on the other hand, significant local minima appear due to the noise. These minima exist in both the timing and the flux methods, but they are deeper in the timing method, while in the flux method, the global minimum is still near the true frequency. We attribute the selection of the wrong frequency in the timing method to the incomplete propagation of errors, a process that assumed a normal distribution of each transit mid-time, whose incorrectness becomes larger and larger as the SNR decreases. Furthermore, we note that a wrong $\omega$ value would lead to wrong $A,\, B$ values that would bias the $\chi^2$ even more, thus making the erroneous global minimum even deeper. Fitting a global flux model does not suffer from this problem: Because, in our experiment, the flux values are distributed normally, and the global minimum in the $\chi^2$ surface is also kept for low-SNR values since the integration of all data points is done appropriately, in the $\chi^2$ definition sense.

\subsection{Results - a Population Analysis}

As explained above, in the transit-by-transit method, the inaccurate error propagation results in an inaccurate estimation of the TTV parameters. We illustrate this in another way by running multiple experiments as described above and checking the distribution of the results among them. For each experiment, we calculate the deviation of the best-fitting parameters from the true value, denoted by $\Delta\omega, \Delta A, \Delta B$, and the error estimation in each parameter, denoted by $\sigma_{\omega},\sigma_A,\sigma_B$. If the estimates distribute normally, then we expect that along multiple experiments with different random noise values, the quantities $\Delta\omega/\sigma_{\omega},\Delta A/\sigma_A,\Delta B/\sigma_B$ will distribute as $\mathcal{N}(0,1)$. 

\begin{figure}
    \includegraphics[width=.9\linewidth]{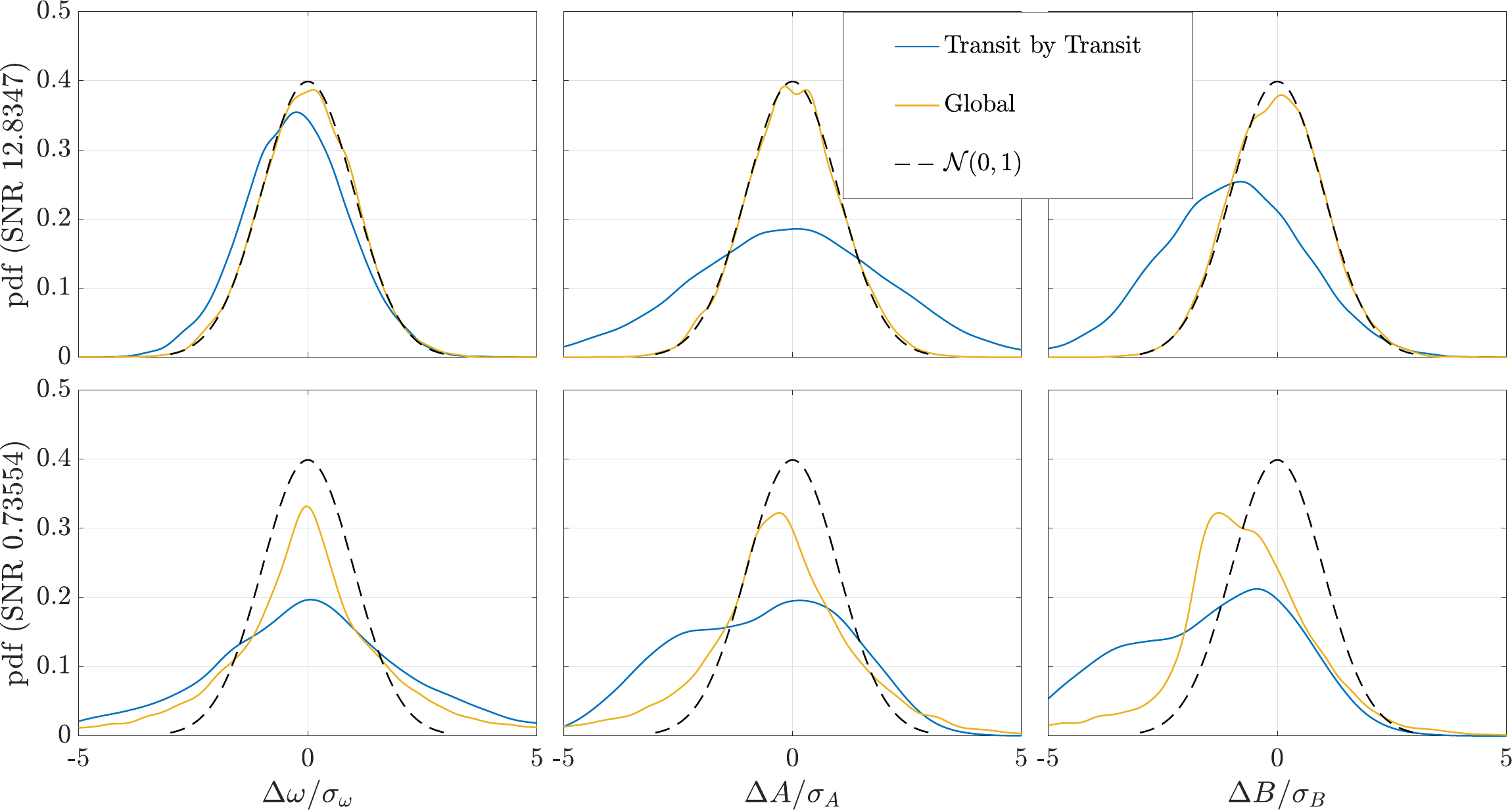}
    \caption{The distribution of the deviation of the parameters estimates from the true values in terms of error estimates, calculated among a set of 15,505 experiments with different noise realizations, shown for two methods: the global fit method and the transit-by-transit method. Top row: the high-SNR case. In this case, the global fit method generates parameter estimates that nicely fit a Gaussian, while the transit-by-transit method generates a broader distribution. The black dashed line indicates a normal distribution with unity standard deviation. Bottom row: the low-SNR case. In this case, both distributions deviate from a Gaussian; the transit-by-transit method generates a much broader distribution, for which the peak is almost erased, while the global fit remains well-centered.}
    \label{fig:SNRVsFlux12}
\end{figure}

In Figure~\ref{fig:SNRVsFlux12} we illustrate the non-Gaussian distribution of the parameters which is attributed to the non-Gaussian PDF of the mid-transit time for each individual event. In this figure, we show the distributions of the quantities $\Delta\omega/\sigma_{\omega},\Delta A/\sigma_A,\Delta B/\sigma_B$ among 15,505 experiments for both the transit-by-transit the global fit method driven by the LM algorithm. We show this for two SNR values. In the high-SNR case, the global fit generates a distribution of the parameters strikingly close to a Gaussian, while the transit-by-transit method generates a broader distribution, clearly indicating a non-Gaussian distribution of the estimated parameters. In the low-SNR case, the distribution deviates from a Gaussian for both methods, albeit to a lesser degree for the global fit method. Interestingly, the transit-by-transit method generates an underestimate of the TTV amplitude. We tested and empirically found that this phenomenon occurs consistently for various values of $A,B$. That is, the total TTV amplitude $\sqrt{A^2+B^2}$ is underestimated, even in cases of high transit SNR. The systematic underestimate of the TTV amplitude would directly translate to an underestimate of planetary mass. This finding could contribute to the previously observed underestimate of TTV masses with respect to RV masses, discussed in \citet{Steffen2016} and \citet{Mills2017}. The modeling bias would also result in an increase in the mass estimates as more data is accumulated, for instance, in the mass estimates of the TRAPPIST-1 system between \citet{GrimmEtAl2018} and \citet{Agol2021}.

In the appendix, we show further plots showing the spread of estimated values along the sample of experiments.

%THE BELOW IS NOT CORRECT. WILL BE REMOVED
% Above, we have shown that for a low SNR value, the $\chi^2$ surface for an individual transit event does not resemble a parabola. Another way to elucidate the non-Gaussian distribution of mid-transit times is to observe the distribution of their deviations from the true mid-transit time. If the likelihood distribution in each event is normal, then the distribution of the deviation of each timing estimate from the true value should distribute like a Gaussian.
% In Figure~\ref{fig:TmidDistribution}, we show this empirical evidence from our experiment. It is apparent that in the high-SNR case, the deviation of the mid-transit times from the actual values distributes normally. In the low-SNR case, this distribution does not resemble a Gaussian with the same standard deviation.

% \clearpage

% \begin{figure}[h]
%     {\includegraphics[width=.9\linewidth]{Figures/SNRVsFlux4.pdf}}
%     {\caption{The cumulative distribution function (CDF) of the errors in mid-transit times for high-SNR (blue) and low-SNR (orange) cases. For comparison, for each CDF we also show a CDF of a Gaussian with a standard deviation taken from the empirical CDF. It is apparent that in the high-SNR case, the observed CDF follow a zero-mean normal distribution closely, while in the low-SNR case, the shape of the distribution deviates from a Gaussian.}
%     \label{fig:TmidDistribution}}
% \end{figure}

\clearpage
\section{Summary}
In this work, we investigated the limits to which a TTV signal can be fitted by estimating the individual mid-transit times and compared them to a method that fits a complete TTV-bearing flux model. We argue that the former approach is limited due to the wrong error propagation from flux to timing. We have shown evidence that this limitation stems from the PDF of each individual transit time, which is not a Gaussian. This trend is correct in all SNR regimes and is enhanced in the case of low single-transit SNR. A (dynamical) fit to the measured TTV signal that assumes a normal distribution of the individual transit times is prone to giving wrong results with underestimated errors, especially in the limit of low SNR. A global fit in the flux fitting method would not suffer from this problem in the high-SNR case and would suffer from this problem to a significantly lesser extent in the low-SNR case. It should be noted, however, that the last stage of timing-based fit, the dynamical stage,  does not rely on an analytic inversion but on a non-linear fitting process and, therefore, would give an answer that might be non-unique or wrong due to convergence to a local minimum. 
As we mentioned above, another issue to be addressed is red noise. In reality, the flux is not white-noised; there is also a red component due to instrumental or astrophysical effects, such as stellar activity or non-transiting planets. This red component would affect both fitting methods; studying how it affects time-fitting versus global flux-fitting is unrelated to the issues investigated here and is not included in this work.

\acknowledgments
This study was supported by the Helen Kimmel Center for Planetary Sciences and the Minerva Center for Life Under Extreme Planetary Conditions \#13599 at the Weizmann Institute of Science.
We thank Barack Zackay and Sahar Shahaf for the helpful discussion on the content of this work.

We thank the reviewer for the useful comments, which improved the quality of this work.

\clearpage
\appendix

\begin{figure}
    \centering
    \includegraphics[width=1\linewidth]{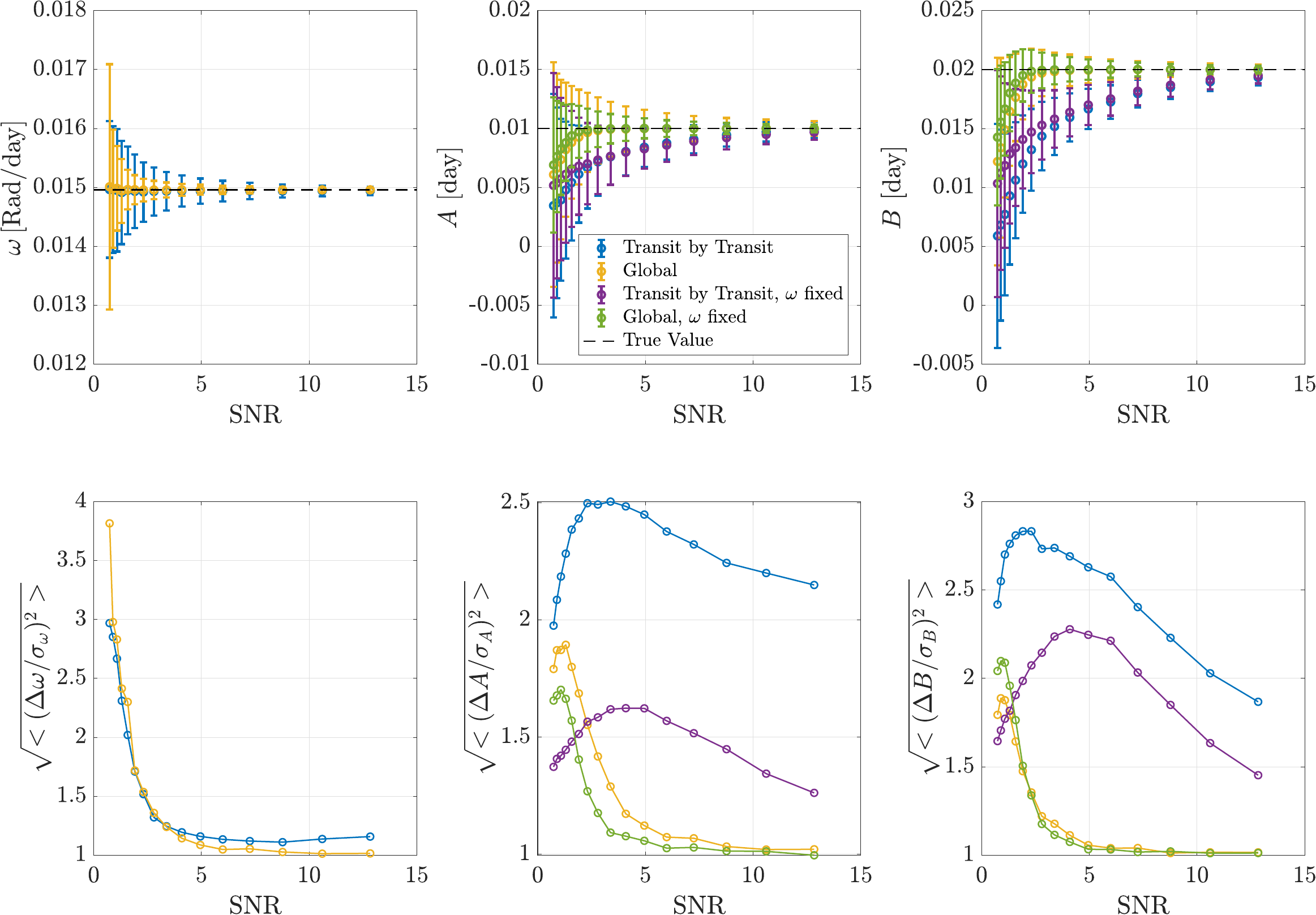}
    \caption{The mean deviation of the parameter estimates from the true value and the normalized deviation in error estimate as a function of transit SNR. Top row: the dots indicate the mean parameter estimate, and the error bars indicate the spread of the parameter estimates. The black dashed line indicates the true parameter value. Bottom row: the square root of the mean square of the deviation-to-error-estimate ratio. Color legend: blue=transit-by-transit, yellow=global flux fit, purple=transit-by-transit with a fixed $\omega$ value, green=global flux fit with a fixed $\omega$ value. See the text for further explanations.}
    \label{fig:SNRVsFlux10}
\end{figure}

In this appendix, we show further evidence from our numerical experiment to elaborate on the difference between the methods applied in this work. We averaged the estimated values over our sample of 15,505 experiments and summed their corresponding error estimates in quadrature. The resulting values and error estimates reflect each method's ability to estimate the parameter's values correctly. The results are shown in figure \ref{fig:SNRVsFlux10}. The top panel in this figure shows the averaged parameter estimates for the utilization of the transit-by-transit method (blue), the global flux method (yellow), and also for these same methods, but when the value of $\omega$ is not fitted to the data, but is fixed to the true value and only $A,B$ are being fit for. The purple curve indicates the results for the transit-by-transit method with a fixed $\omega$ value, and the green curve shows the results for the global flux method with a fixed $\omega$ value. The black dashed lines indicate the true parameter values.

The results show an apparent rise in the normalized deviation from the true value of the TTV frequency as the SNR reaches a value of 2-3. This is true for both methods. The values of $\omega$ themselves distribute well around the true value. For ${\rm SNR}\gtrsim 3$, both methods yield estimates of $\omega$ that average to $\approx\,1$, meaning that the distribution is close to a Gaussian, with the global flux fit method slightly closer to unity than the transit-by-transit method. An inspection of the corresponding plots for the parameters $A,B$ shows that for ${\rm SNR}\gtrsim 3$, the global flux method outperforms the transit-by-transit method in terms of the correctness of both the values and the error estimates of the fitted parameters. The top row panels show that the values are distributed around the true value in the flux-fitting method. The lower panel shows that the normalized deviation of $A,B$ from the true value is very close to unity in the flux-fitting method and significantly larger than unity in the transit-by-transit method. Having an exact knowledge of the TTV frequency improves the estimates in the transit-by-transit method but still does not bring it to the level of normal distribution. We stress that TTV frequency is related to the orbital periods of the planets, while the amplitude and phase are the parameters that are directly linked to the planetary masses and eccentricities \citep[][and others]{LithwickXieWu2012,HaddenLithwick2014,Judkovsky2022a}.

\clearpage
\bibliography{Mybib}

\end{document}